\documentstyle[epsf,aps,preprint,floats]{revtex}

\begin{document}

\preprint{}

\title{A Bound on the Flux of Magnetic Monopoles from Catalysis of Nucleon
Decay in White Dwarfs}
\author{Katherine Freese and Eleonora Krasteva}
\address{Department of Physics, University of Michigan, Ann Arbor,
		Michigan 48109-1120}

\date{submitted to Physical Review D, April 15, 1998}

\maketitle

\begin{abstract}
Catalysis of nucleon decay in white dwarfs is used to constrain the abundance
of magnetic monopoles arising from Grand Unified Theories. Recent discoveries
of the dimmest white dwarf ever observed, WD 1136-286 with
$L = 10^{-4.94} L_{\odot}$, place limits on the monopole flux.
An abundance of monopoles greater than the new bound
would heat this star to a
luminosity higher than what is observed. The new bound is
$(F/$cm $^{-2}$ s$^{-1}$ sr$^{-1}$) 
$(\sigma \upsilon/10^{-28} {\rm cm}^2) < 1.3 \times 10^{-20} (\upsilon/
10^{-3}c)^2$, where $\upsilon $ is the
monopole velocity. The limit is improved by including the monopoles 
captured by the main-sequence progenitor of the white dwarf:
$(F/$cm $^{-2}$ s$^{-1}$ sr$^{-1}$ ) 
$(\sigma \upsilon /10^{-28} {\rm cm}^2) < 3.5(26) \times 10^{-21}$
for $10^{17}$ ($10^{16}$) GeV monopoles.

We also note that the dependence on monopole mass of flux bounds
due to catalysis in neutron stars with main sequence accretion
has previously been calculated incorrectly (previously
the bound has been stated as 
$F (\sigma \upsilon/10^{-28} {\rm cm}^2) <  10^{-28} $ 
cm $^{-2}$ s$^{-1}$ sr$^{-1}$).  We show that
the correct bounds are somewhat weaker for monopole mass
other than $10^{17}$ GeV.

\end{abstract}

\newpage

\section{Introduction}
The question of whether or not magnetic monopoles exist has intrigued 
theorists and experimentalists for a long time \cite{[1]}. In 1974, t'Hooft 
\cite{[2]} and Polyakov \cite{[3]} independently showed that magnetic 
monopoles always appear as 
stable topological entities in any Grand Unified theory (GUT) 
that breaks down to
electromagnetism. Hence, if Grand Unified theories are shown to be correct,
monopoles of mass in the range $10^{15}$ - $10^{19}$ GeV should exist. Rubakov
\cite{[4]} and Callan \cite{[5]} calculated that these monopoles catalyze 
nucleon decay with a cross section characteristic of strong interactions,
 $\sigma \upsilon \approx 10^{-28}$ cm$^2$. 

The abundance of these monopoles is an open question. The Kibble
mechanism predicts roughly one monopole per horizon volume at the time
of the Grand Unified phase transition. However, this estimate provides
a severe overabundance of the number of monopoles: monopoles 
overclose the Universe by many orders of magnitude. Instead  an inflationary 
epoch \cite{[6]} may reduce their density in the Universe. Then the  
present abundance is difficult to estimate. A clue for experimentalists 
about what monopole flux to expect can be provided by astrophysics.

The Parker bound \cite{[7]} 
on the flux of monopoles was obtained by requiring survival 
of $\mu$G magnetic fields observed in our Galaxy and gave $F \leq 10^{-16}$
cm$^{-2}$ sr$^{-1}$ sec$^{-1}$.  Subsequent improvements 
on this work 
include consideration of the monopole velocities \cite{[8]} 
due to acceleration by the galactic magnetic field. Another
improvement is the extended Parker bound, which required survival of a 
smaller seed magnetic field in the early period of the Galaxy \cite{[9]}:
 $F \leq 1.2 \times 
10^{-16} (\frac {m}{10^{17}GeV})$ cm$^{-2}$ s$^{-1}$ sr$^{-1}$.

Another class of methods for determination of the monopole flux is based on 
the hypothesis that GUT monopoles give rise to the catalysis of 
nucleon decay. The basic idea is that monopoles 
traveling through the Galaxy lose enough energy to be captured in an object 
(e.q. white dwarfs, neutron stars, etc.) where they subsequently catalyze 
nucleon decay. The energy produced by the nucleon decay heats up the object 
and results in a flux of photons from the surface of the object. 
One can then compare this predicted luminosity with what is actually observed.
One must ensure that the monopoles would not make the object brighter than 
what is seen. The coolest star (or other object) seen provides the tightest 
limit on the monopole flux. If there were more monopoles than allowed by the 
bound, then the dimmest star observed could not exist.

Several authors have carried out this kind of analysis in neutron stars 
\cite{[10]}, 
nearby pulsar and white dwarfs. 
The strongest bound was obtained from consideration
of the catalysis process in PSR 1929+10, an old pulsar \cite{[11]}. From
this pulsar, the bound on the product of monopole flux
times cross section for catalysis is
$(F/$cm$^{-2}$sr$^{-1}$sec$^{-1})(\sigma\upsilon / 10^{-28}$ cm$^{2}) \leq 7 
\times 10^{-22}$. If one includes the monopoles captured by the main sequence
progenitor of the white dwarf, this bound becomes even tighter \cite{[12]},
\hfill\break
$(F/$cm$^{-2}$ sr$^{-1}$ sec$^{-1})(\sigma\upsilon 
/ 10^{-28}$ cm$^2$ $) \leq 10^{-28}$.
The consideration of monopole dynamics inside superconducting
neutron-star cores leads to a bound $5 \times 10^{-24} \tau_{10}^{-2}$
cm$^{-2}$ sr$^{-1}$ s$^{-1}$ \cite{[13]}, where $\tau_{10}$ is the age (in 
$10^{10}$ years) of the pulsar's present magnetic field.

As neutron stars are the densest astrophysical objects observed,
they give rise to the tightest catalysis bounds. However, there is a certain
amount of uncertainty due to the fact that the interiors of neutron stars are 
not well understood. For example, neutron stars can have very large magnetic 
fields $\sim 10^{12}$G of unknown topology, and the motion of magnetic 
monopoles inside the neutron star would undoubtedly be affected by these 
magnetic fields.  In addition neutron star interiors may
contain pion condensates, again with uncertain effects on the monopoles.
Because of the uncertainties with neutron star interiors, we turn to the next
densest astrophysical objects in the Universe, white dwarfs. 
These stellar remnants are far better
understood. The flux limits obtained from consideration of the catalysis
process in white dwarfs are therefore important. Previously Freese 
\cite{[14]} considered monopole catalyzed nucleon decay in white dwarfs.
By comparing with the lowest luminosity white dwarf that 
had been seen at that time, she obtained a limit

\begin{equation}
(F/{\rm cm}^{-2}{\rm s}^{-1}{\rm sr}^{-1})
(\sigma\upsilon /10^{-28}{\rm cm}^2) \leq 2\times 10^{-18}.
\end{equation}

The present work is motivated by new observational data of cool white dwarfs
\cite{[15]}. 
In particular, Bergeron, Ruiz, and Leggett found a
white dwarf 1136-286 (ESO 439-26)
with luminosity $10^{-4.94}L_{\odot}$; this is the dimmest white 
dwarf observed to date.  We use the measured luminosities of old white dwarfs 
to constrain the radiation due to monopole-catalyzed nucleon decay and thus 
to obtain an upper limit to the average flux of monopoles in the Galaxy. Since
a white dwarf with luminosity $10^{-4.94}L_{\odot}$ is observed
today, we know that the monopole-induced contribution to the white dwarf
luminosity cannot exceed this value. These new data improve the limit on the 
monopole abundance due to catalysis in white dwarfs \cite{[14]} by roughly
two orders of magnitude. Of course, as dimmer white dwarfs 
are found, the bound will continue to get more restrictive.

A monopole flux saturating this bound would keep the white dwarfs at 
luminosities at least this great and would lead to the prediction that no 
cooler white dwarfs will be found. As we will discuss, if it were indeed true
that monopoles are keeping dwarfs hot, one would expect a different dependence
of white dwarf luminosity on mass than expected in the standard model. 

We shall explicitly indicate the dependence of our 
results on various parameters. We will parametrize the properties of the white
dwarf in terms of typical values from observations: for the mass, 
$M = M_{0.6} 0.6 M_{\odot}$, for the radius $R = R_{9} 9\times 10^8$cm, and 
for the average density 
$\bar \rho = 4\times 10^5$ g cm$^{-3} M_{0.6} R_{9}^{-3}$. The central density
is about an order of magnitude higher, $\rho_{c} = 3 \times 10^6$ g cm$^{-3} 
M_{0.6} R_{9}^{-3}$. Rubakov \cite{[4]} estimated the product of cross section 
for catalysis and relative velocity $\upsilon$ of the monopole and nucleon to
be constant: $\sigma \upsilon = \sigma_{0} = 10^{-28} ({\sigma\upsilon}_{-28})$
cm$^2$. (Throughout, we take $\hbar  = k_{B} = c = 1$.) For the thermal nucleon
velocities expected inside a carbon and oxygen white dwarfs, $\upsilon \approx 
10^{-3}c$, suppression effects may reduce the cross section by a factor of
$10^{-2} s_{-2}$ \cite{[16]}, and so we include this factor. 
In white dwarfs made of helium the suppression effects would be less effective
$(s_{-2} = 10)$, and all the monopole flux bounds would be an order of 
magnitude stronger.

\section{Estimation of the monopole flux}

As we noted in the introduction, monopole-catalyzed nucleon decay caused by the
monopoles captured in white dwarfs can provide an additional internal heat 
source for
the star. Our evaluation of the monopole flux is based on the observed
luminosity of the white dwarfs and estimation of the number of monopoles
trapped inside the stars. As a monopole passes through a white dwarf, it loses
energy and is captured. Electronic interactions are considered to be the 
primary source of energy loss for the monopoles, with \cite{[17]},
\cite{[18]},
\begin{equation}
\frac{dE}{dx} = \frac{2\pi n_{e}(eg)^2 \beta}{k_F} \left( \ln 
\frac{1}{Z_{min}} - \frac{1}{2} \right) \approx 100\rho\beta 
 {\rm GeV/cm}, 
\end{equation}
 where $n_e$ is the
electron density, the Fermi momentum of the electrons $k_F \approx 0.1$MeV, 
$Z_{min} = 2k_{F}\lambda/h$, $\lambda$ is the mean free path of the 
electron, $\rho$ is the density of the white dwarf (in ${\rm g cm}^{-3}$), and 
$\beta$ is the velocity of the monopole as it passes through the white dwarf.

As was shown in \cite{[14]}, a white dwarf accumulates almost all the
monopoles with $m \leq 10^{20}$ GeV incident upon it. The number of monopoles
captured by a white dwarf exposed to a monopole flux  $F($cm$^{-2}$ $s^{-1}$
sr$^ {-1})$ for a time $\tau = \tau_{10} 10^{10}$ yr is given by

\begin{equation}
 N_{M} \approx FA \tau(\pi sr) \approx 2.0 \times 10^{39} a_{1} F
\end{equation}
 where 
$A = 4 \pi R^{2} [1 + (2GM/ R \upsilon_M^2)]$ is 
the capture area, $\upsilon_M = \upsilon_{-3}  10^{-3}c$ is the monopole 
velocity and
$a_1 = \tau_{10} R_{9} M_{0.6} v_{-3}^{-2}$.

Once captured, the monopoles sink to the center of the white dwarf. In
calculating the luminosity from catalyzed nucleon decay, 
we use the central density
of the white dwarf. We are justified in doing this since the time for the
monopole to fall (from rest) into the center is $\approx 1000s$, as  has been 
calculated \cite{[14]} by treating the 
motion of the monopoles as a harmonic  oscillator
with a $dE/dx$ damping term. We find the luminosity from catalyzed nucleon 
decay per monopole:

\begin{equation}
L_{1} = \rho_{c}\sigma\upsilon = 8.1 \times 10^7 {\rm erg s^{-1} }
(\sigma\upsilon)_{-28} s_{-2} M_{0.6} R_{9}^{-3}.
\end{equation}

Then the total luminosity of a white dwarf due to a monopole-catalyzed nucleon
decay is

\begin{equation}
L_{mon} = N_{M} L_{1} = 1.6\times 10^{47} a_{2} F {\rm erg s}^{-1},
\end{equation}
where $a_{2} = \tau_{10} R_{9}^{-2} M_{0.6}^{2} (\sigma\upsilon)_{-28} s_{-2} v_{-3}^{-2}$.

From the Stefan-Boltzman law, $L_{mon} = \sigma_{BB} 4 \pi R^{2} T_{eff}^{4}$,
where $\sigma_{BB}$ is the Stefan-Boltzman constant, 
we can find the blackbody temperature corresponding to this
luminosity,

\begin{equation}
T_{eff}= 1.3 \times 10^{8} {\rm K} (F a_{2} R_{9}^{-2})^{1 \over 4}.
\end{equation}

White dwarfs cool as they age. The cooling time is a function of the white 
dwarf mass and composition. For white dwarf 1136-286,
we use the mass and composition provided by the observers \cite{[15]}.  
The observed energy distributions are obtained from a combination of
both optical BVRI and infrared HJK photometric data and used to 
derive both the effective temperature and the atmospheric
composition of the star.  This white dwarf is seen
to have a Helium atmosphere.  Stellar masses were also obtained with
trigonometric parallax. Bergeron, Ruiz, and Leggett \cite{[15]} derive
$M=1.2 M_{\odot}$ for WD 1136-286. Then from
measurements of the luminosity and $T_{eff}$,
Eqn. (6) implies that the radius is $R \approx 3.9\times 10^8$cm. 

We also use two different cooling models. First we use the white dwarf cooling
theory from the calculations of Segretain et al. \cite{[19]},
as communicated by G. Chabrier. The Segretain et al. \cite{[19]} model 
accounts for gravitational energy 
release due to carbon-oxygen differentiation at crystallization. This
treatment of crystallization yields significantly longer white dwarf cooling
times, which in turn imply an older age for any particular white dwarf.
These white dwarf models correspond to a mass sequence of initially
unstratified white dwarfs composed of equal parts carbon and oxygen, with 
helium atmospheres. With these models, the age of white dwarf 1136-286
is 9.63 Gyr.  For comparison we also use the cooling curves of Wood 
\cite{[20]} which do not include chemical fractionation. Chemical fractionation
provides an additional source of energy to be radiated away; thus models that
lack it cool faster. With the Wood cooling models, the ages of white 
dwarfs are somewhat younger. Hence these models give younger white 
dwarfs that accumulate somewhat fewer monopoles and provide somewhat less
restrictive bounds. With the Wood cooling curve, the age
of white dwarf 1136-286 is 6.47 Gyr.
To illustrate the uncertainty we provide flux bounds using both possible
ages, but note that the discrepancy is not very great.

The cooling models discussed above do not yet have an additional heat source 
due to monopoles. If white dwarfs have indeed been accumulating monopoles, then
the monopole contribution to the luminosity increases linearly in time, and 
monopole catalyzed nucleon decay will eventually become the dominant source
of luminosity. The minimum value of the total luminosity must be at least as
low as $10^{-4.94} L_{\odot}$, since white dwarf WD 1136-286
with this luminosity has been observed to exist. Using the 
mass and radius discussed previously for this white dwarf, we then
find from Equations (3-5) that
\begin{equation}
N_{M} \leq 2.2 \times 10^{19}(\sigma \upsilon)_{-28}^{-1} s_{-2}^{-1}.
\end{equation}

With the cooling curves of Segretain et al., which 
include the effects of chemical fractionation, the age for this 
particular white dwarf WD 1136-286 is given to be 9.63 Gyr 
as mentioned above.  We find a flux bound 

\begin {equation}
F \leq 1.3 \times 10^{-20} (\sigma\upsilon)^{-1}_{-28} s^{-1}_{-2}
{\upsilon}^2_{-3} {\rm cm}^{-2}{\rm s}^{-1} {\rm sr}^{-1} \, .
\end{equation}

With the Wood \cite{[20]} cooling curves, the age of the white dwarf 
is 6.47 Gyr as mentioned above.  Then
equation (6) corresponds to a flux bound

\begin{equation}
F \leq 1.9 \times 10^{-20} (\sigma\upsilon)^{-1}_{-28} 
s^{-1}_{-2} {\upsilon}^2_{-3} {\rm cm}^{-2}{\rm s}^{-1}{\rm sr}^{-1} \, .
\end{equation}

This bound using the Wood cooling curves is less restrictive
than the one obtained using the Segretain et al. cooling curves.
Hence, to be conservative, in Figure 1 we plot the flux bound of Eq. (9).
Note that 
the monopole velocities  have been determined as a function of monopole 
mass by the following equation: $\upsilon_{M} \approx 3 \times
10^{-3} c(10^{16} {\rm GeV/m})^{1/2}$ for monopole mass
$m<10^{17}$ GeV and $\upsilon_{M} \approx 10^{-3} c$
for monopoles with mass greater or equal to $10^{17}$ GeV \cite{[8]}. 
Thus the flux bound is
flat for monopole masses greater than $10^{17}$ GeV and drops as 
m$^{-1}$ for smaller masses.  This behavior can be seen in Figure 1.

If the monopole flux saturates the bound in equations (8) and (9), 
the heat release due
 to monopole-catalyzed nucleon decay would explain the dearth of white dwarfs 
with luminosity $\leq 10^{-5} L_{\odot}$. That is, monopoles may be keeping 
white dwarfs hot. 
Note that the white dwarf luminosity due to monopole
catalyzed nucleon decay scales as $L_{mon} \propto \tau_{10}
M_{0.6}^2$.  If the luminosity of the coolest objects
we see today is in fact due primarily to the contribution
from monopoles, then one would in principle be able to see this
dependence on white dwarf mass.  However, one would need
to be able to independently measure the white dwarf luminosity,
mass, and age in order to test this hypothesis.

In an earlier paper Freese \cite{[14]} checked that 
the presence of monopoles did 
not drastically affect the properties of the white dwarf in any way. A usual 
white dwarf is an isothermal, electron degenerate object surrounded by a very 
thin radiative envelope. The primary mechanism of heat transfer through the 
body of the star is conduction. In the 
presence of monopoles the white dwarf remains essentially isothermal, with a 
radiative envelope, conductive main body, and convective core so that one may 
conclude that monopoles have a negligible effect on the overall structure of 
white dwarfs (for more details see \cite{[14]}).

{\it Monopole/antimonopole annihilation:}
As discussed in Freese \cite{[14]}, monopole-antimonopole annihilation has no 
effect on the flux bound obtained in equations (8) and (9). There it was shown
that, if the above flux bounds are satisfied, the number of monopoles 
accumulating inside the white dwarf never reaches a sufficient abundance for 
annihilation to become significant.

\section{Tighter Bounds Obtained if Monopoles Captured by
the Main Sequence Progenitor are Included:}

During its main sequence period the progenitor of the white dwarf may
also have captured a significant number of monopoles. These additional
monopoles will lead to an even tighter bound on the monopole flux.
In order to estimate the number of monopoles captured by the main
sequence progenitor, we must determine its mass. Unfortunately, the 
transformation from main sequence mass to white dwarf mass is somewhat 
uncertain, as discussed by Wood \cite{[20]}. Here we use
$M_{WD} = A_x {\rm exp} (B_x M)$, where $A_x = 0.49$ and $B_x = 0.095$.
For  white dwarf 1136-286 with mass $M_{WD} = 1.2 M_{\odot}$, we find 
progenitor mass $M = 9.4 M_{\odot}$.

As a monopole passes through a MS star, it loses energy. If it loses all 
its initial kinetic energy (i.e. its energy infinitely far from the star),
it is captured by the star. Since the energy loss increases with decreasing 
impact parameter, the number of monopoles captured by a MS star exposed
to a monopole flux F for a time $\tau = \tau_{MS}$  is just the number
incident upon the star with surface impact parameter less than some
critical value, $b_{crit}$:
\begin{equation}
N_{M} = (4\pi b_{crit}^2) (\pi sr) [1 + (\frac{\upsilon_{esc}}
{\upsilon_{\infty}})^2 ] F \tau_{MS},
\end{equation}
where $\upsilon_{\infty}$ is the monopole velocity far from the star
and $\upsilon_{esc} = (2GM/R)^{1/2}$ is the escape velocity from the star.
Frieman, Freese, and Turner \cite{[12]} previously calculated numerically
the critical impact parameter for capture. We use those results 
here\footnote{Recent calculations of Ahlen and Tarle \cite{[21]} indicate
that equation (2) for the energy loss of monopoles in main sequence
stars must be increased by a factor of 2.  Thus $b_{crit}$ should
be somewhat larger than what was calculated by Frieman, Freese, and
Turner \cite{[12]}.  
For example, for the case where $\beta = \beta_{esc} = 3.3 \times 10^{-3}$
(for a 9 $M_\odot$ star), rough analytic estimates indicate
that the modified value of $b_{crit}$ is roughly given by
$\bigl({b_{crit} \over R }\bigr)_{\rm new}^2 = 0.058 + 
\bigl({b_{crit} \over R }\bigr)_{\rm old}^2 $. 
As the calculations of \cite{[12]} are numerical and the difference
is very small, we will continue to use the prior results of \cite{[12]}.  
In fact, if one were to use the newer value of $b_{crit}$, the flux
bounds would be somewhat tighter.}.
Given the value of $b_{crit}$, we can substitute it into the previous 
equation (11) to obtain the number of monopoles captured by the main 
sequence progenitor. The sum of monopoles captured by the progenitor 
plus those captured by the white dwarf cannot exceed the maximum number
allowed in Equation (7), so that we obtain a new flux bound. Again,
we have taken monopole velocities to be $\upsilon_{M} \approx 3 \times
10^{-3} c(10^{16} {\rm GeV/m})^{1/2}$ for monopole mass
$m<10^{17}$ GeV and $\upsilon_{M} \approx 10^{-3} c$
for monopoles with mass greater or equal to $10^{17}$ GeV \cite{[8]}.
The inclusion of monopoles captured by the progenitor of the white dwarf
results in a bound on monopole flux that
is another order of magnitude lower. We have 
evaluated the new bound for both white dwarf cooling models. 
In Table 1, we have recorded, as a function of monopole mass,
the value of $b_{crit}/R$, the number of monopoles captured by
the progenitor and white dwarf, and the resultant flux bound. 
The flux bound is plotted in Figure 1.

The flux bound is most restrictive for $m \sim 10^{17}$ GeV.  
One can understand the
reason for the weaker flux bounds for masses greater
and less than $10^{17}$ GeV as follows:  For masses less 
than this, it is the increasing monopole velocity as a function
of decreasing monopole mass that drives
the flux bound to become weaker for smaller mass.
For masses $m> 10^{16}$ GeV, the factor $[1 + (\frac{\upsilon_{esc}}
{\upsilon_{\infty}})^2 ]$ in Eq. (10) is dominated by the 
second term $(\frac{\upsilon_{esc}} {\upsilon_{\infty}})^2 $.
As the monopole mass decreases below $10^{17}$ GeV,
the monopole velocity increases,  the term
$(\frac{\upsilon_{esc}} {\upsilon_{\infty}})^2 $ decreases,
so that in Eq. (10) the monopole flux can increase
and still maintain the same number of monopoles in the star
and hence the same luminosity.  Eventually, when
the mass drops to 
$m \leq 10^{16}$ GeV, the first term, $1$, starts to dominate in
the factor $[1 + (\frac{\upsilon_{esc}} {\upsilon_{\infty}})^2 ]$,
so that the monopole velocity becomes unimportant
and the curve becomes more and more flat with decreasing monopole mass.
This behavior can be seen in Figure 1.
For monopole masses greater than $10^{17}$ GeV, the flux bound
also becomes weaker, this time as a function of increasing
mass.  The reason for this is as follows.  
These monopoles all move with the virial velocity
of the Galaxy $\sim 10^{-3}c$.  The heavier the monopole,
the harder it is to stop.  Hence $b_{crit}$ becomes smaller for
heavier masses.  Thus a larger monopole flux can be accomodated
in Eq. (10) to still obtain the same number of monopoles in the star.

\section{Neutron Stars with Main Sequence Accretion:}

We also note that the dependence on monopole mass of flux bounds
due to catalysis in neutron stars with main sequence accretion
has previously been calculated incorrectly.  In the past
the bound due to catalysis in PSR 1929+10 with main sequence
accretion has been stated as \cite{[11]}
$F (\sigma \upsilon/10^{-28} {\rm cm}^2) <  10^{-28} $ 
cm $^{-2}$ s$^{-1}$ sr$^{-1}$.  
As discussed in the previous paragraph, the velocity dependence
of monopoles of different masses determines the shape of the
curve of flux bounds as a function of monopole mass.  As can be
seen in Table 1 and Figure 1,
the correct bounds are somewhat weaker for monopole mass
other than $10^{17}$ GeV because of the faster velocities of monopoles
with smaller masses and the lower critical impact parameter
for monopoles with larger masses.  In obtaining the numbers, we have assumed
a main sequence progenitor of 9 $M_\odot$.  Then the
number of monopoles captured by the main sequence progenitor
of the neutron star is the same as the number of monopoles
captured by the white dwarf considered in this paper.

\section{Conclusion}

Figure 1 
shows a plot of several monopole bounds: the Parker bound, the extended
Parker bound, neutron star bounds, and the new white dwarf bound with
and without main sequence capture.  In the plots we
have used the Wood cooling curves to be conservative.
We have found that consideration
of newly observed white dwarf 1136-286 with luminosity
$10^{-4.94} L_\odot$ and with new calculations
of white dwarf cooling curves leads to a bound on the monopole flux that
is two orders of magnitude lower than previous bounds
due to catalysis in white dwarfs.  The new bound is
$F (\sigma \upsilon/10^{-28} {\rm cm}^2) < 1.3 (1.9) \times 10^{-20} 
(v/10^{-3}c)^2$ 
cm $^{-2}$ s$^{-1}$ sr$^{-1}$ for the
Segretain \cite{[19]} (Wood \cite{[20]}) cooling 
curves respectively, where $\upsilon $ is the
monopole velocity. The limit is improved by including the monopoles 
captured by the main-sequence progenitor of the white dwarf:
$F (\sigma \upsilon /10^{-28} {\rm cm}^2) < 3.5(26) \times 10^{-21}$
cm$^{-2}$ s$^{-1}$ sr$^{-1}$ 
for $10^{17}$ ($10^{16}$) GeV monopoles with $g=g_D$.   
Flux bounds for other monopole masses and parameters are given in Table 1.
If cooler white dwarfs are discovered, 
a stricter bound on the monopole flux will result.

We also showed that the dependence on monopole mass of flux bounds
due to catalysis in neutron stars with main sequence accretion
has previously been calculated incorrectly.  Previously
the bound due to catalysis in PSR 1929+10 with main sequence
accretion has been stated as \cite{[11]}
$F (\sigma \upsilon/10^{-28} {\rm cm}^2) <  10^{-28} $ 
cm $^{-2}$ s$^{-1}$ sr$^{-1}$.  
Instead, as can be
seen in Table 1 and Figure 1,
the correct bounds are somewhat weaker for monopole mass
other than $10^{17}$ GeV.

{\bf Figure Caption}
Bounds on the monopole flux as a function of monopole mass.
The Parker bound \cite{[7]} due to survival of the galactic 
magnetic field is plotted,
as is the extended Parker bound \cite{[9]} due to survival of the magnetic
field early in the history of the Galaxy.  
Mass density limits ($\Omega h^2 <1$) are plotted for a uniform density
of monopoles in the universe.
The bounds due to catalysis
in white dwarf WD1136-286 as discussed in this paper are plotted;
the plots assume the cooling curves of Wood \cite{[20]},
and are very similar to those
obtained using cooling curves of Segretain ${\it et al.}$
In addition, the bounds from this white dwarf with
main sequence accretion (WD/MS) are plotted for $g=g_D$ (solid line) and
for $g=2 g_D$ (dotted line).  The bounds due to calaysis in 
neutron star PSR 1929+10 are plotted, as are bounds due to this
neutron star with main sequence accretion.  Again the solid line
is for $g=g_D$ and the dotted line is for $g=2g_D$.  Note that the
neutron star bounds with main sequence accretion have dependence
on the monopole mass.

\vfill\eject
\begin{center}
\begin{tabular}{|c|c|c|c|c|c|c|c|}\hline
&&&&&&$WD {\rm w}/MS$&$NS {\rm w}/MS$\\ \cline{7-8}
$M_m$ (GeV)&$\beta$&${b_{\rm crit}\over R}$&$g/g_D$
&$N_{MS}/10^{38}F$&$N_{WD}/10^{38}F$&$F(\sigma\upsilon)_{-28}$&$
F(\sigma\upsilon)_{-28}$ 
\\[5pt] \hline
\lower1.5ex\hbox{$10^{15}$}&\lower1.5ex\hbox{$10^{-2}$}
&{0.4}&{1}&2.5 &\lower1.5ex\hbox{0.17}
&$8.2\times 10^{-20}$& $6.2\times 10^{-27}$ \\ \cline{3-5}\cline{7-8}
&&0.56&{2}&4.9 && $4.3\times 10^{-20}$& $3.2\times 10^{-27}$ 
\\\hline
\lower1.5ex\hbox{$10^{16}$}&\lower1.5ex\hbox{$3\times
10^{-3}$}&0.48&1&7.4&\lower1.5ex\hbox{1.8}&$2.4\times 10^{-20}$&
$2.1\times 10^{-27}$ \\ \cline{3-5}\cline{7-8} 
&&0.62&2&12.3 && $1.6\times 10^{-20}$& $1.3\times 10^{-27}$ 
\\\hline
\lower1.5ex\hbox{$10^{17}$}&\lower1.5ex\hbox{$10^{-3}$}&0.54&1&52&\lower1.5ex\hbox{17}& $3.2\times 10^{-21}$& $3.0\times 10^{-28}$
 \\
\cline{3-5}\cline{7-8} &&0.68&2&82 &&$2.2 \times
10^{-21}$& $1.9\times 10^{-28}$ \\ \hline
\lower1.5ex\hbox{$10^{18}$}&\lower1.5ex\hbox{$10^{-3}$}&---&1&---&
\lower1.5ex\hbox{17 }&---&---\\
\cline{3-5}\cline{7-8}&&0.24&2&10 &&$8.1 \times
10^{-21}$& $1.6\times 10^{-27}$ \\\hline
\end{tabular}
\end{center}
\bigskip
\begin{center}
\parbox{6in}{\small Table 1: 
Bounds on the flux $F$ of magnetic monopoles in cm$^{-2}$s$^{-1}$sr$^{-1}$.
Monopoles captured by white dwarfs (WD) or neutron stars (NS)
and their main sequence (MS) progenitors are included.  
The white dwarf cooling time is taken to be $\tau = 9.63$ Gyr.  $M_m$ is the
monopole mass in GeV, $\beta$ is the monopole velocity,
${b_{\rm crit} \over
R}$ is the ratio of the critical impact parameter for a monopole in units
of the radius of the main sequence star, and the monopole charge is $g=
69 e (g/g_D)$ in units of the Dirac charge $g_D$.  The number of
monopoles captured by the MS progenitor and by the white dwarf
are $N_{MS}$ and $N_{WD}$ respectively.  The second to last column
is the flux bound due to catalysis in WD 1136-286 (with MS monopoles
included).  The last column
is the (corrected) flux bound due to catalysis in neutron star
PSR 1929+10 (with MS monopoles included).}
\end{center}
\vspace{10mm}
\begin{center}
\begin{tabular}{|c|c|c|c|c|c|c|}\hline
$M_m$ (GeV)&$\beta$&${b_{\rm crit}\over
R}$&$g/g_D$&$N_{MS}/10^{38}F$&$N_{WD}/10^{38}F$&$F(\sigma\upsilon)_{-28}/
{\rm cm}^{-2} {\rm s}^{-1} {\rm
sr}^{-1}$ \\[5pt] \hline
\lower1.5ex\hbox{$10^{15}$}&\lower1.5ex\hbox{$10^{-2}$}
&{0.4}&{1}&2.5 &\lower1.5ex\hbox{0.11}
&$8.4\times 10^{-20}$  \\ \cline{3-5}\cline{7-7}
&&0.56&2&4.9 && $4.4\times 10^{-20}$ 
\\\hline
\lower1.5ex\hbox{$10^{16}$}&\lower1.5ex\hbox{$3\times
10^{-3}$}&0.48&1&7.4 &\lower1.5ex\hbox{1.2}&$2.6\times 10^{-20}$ \\ \cline{3-5}\cline{7-7} 
&&0.62&2&12.3 && $1.6\times 10^{-20}$
\\\hline
\lower1.5ex\hbox{$10^{17}$}&\lower1.5ex\hbox{$10^{-3}$}&0.54&1&52 &\lower1.5ex\hbox{11 }& $3.5\times 10^{-21}$
 \\
\cline{3-5}\cline{7-7} &&0.68&2&82 &&$2.4 \times
10^{-21}$  \\ \hline
\lower1.5ex\hbox{$10^{18}$}&\lower1.5ex\hbox{$10^{-3}$}&---&1&---&
\lower1.5ex\hbox{11 }&--- \\
\cline{3-5}\cline{7-7}&&0.24&2&10 &&$1.0 \times
10^{-20}$  \\\hline
\end{tabular}
\end{center}
\bigskip
\begin{center}
{\small Table 2:
Same as table 1 for white dwarfs, but for cooling time $\tau = 6.47$ Gyr.}
\end{center}

{\bf Acknowledgements:} We thank J. Allyn Smith, D. Graff, G. Laughlin,
G. Tarle, and V.D. Ivanov for helpful conversations.
We acknowledge support from the DOE at the University of Michigan.

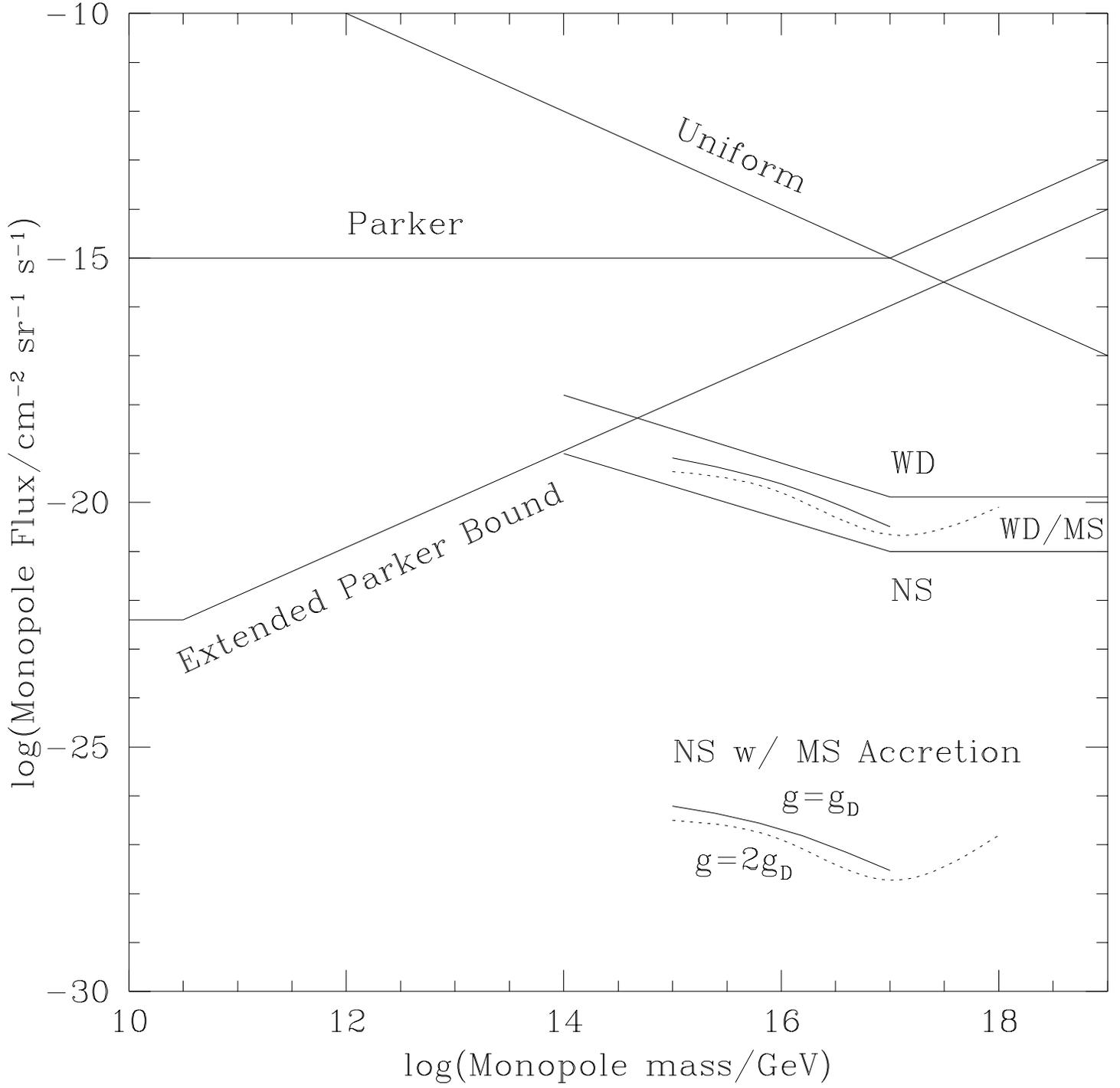
\end{document}